# Association of normalization, non-differentially expressed genes and data source with machine learning performance in intra-dataset or cross-dataset modelling of transcriptomic and clinical data

Fei Deng[1], Lanjing Zhang[1, 2, 3, *]

[1]Department of Chemical Biology, Ernest Mario School of Pharmacy, Rutgers University, Piscataway, NJ, USA.

[2] Department of Pathology, Princeton Medical Center, Plainsboro, NJ, USA.

[3] Rutgers Cancer Institute of New Jersey, New Brunswick, NJ, USA.

## Abstract:

Cross-dataset testing is critical for examining machine learning (ML) model's performance. However, most studies on modelling transcriptomic and clinical data only conducted intra-dataset testing. It is also unclear whether normalization and non-differentially expressed genes (NDEG) can improve cross-dataset modeling performance of ML. We thus aim to understand whether normalization, NDEG and data source are associated with performance of ML in cross-dataset testing. The transcriptomic and clinical data shared by the lung adenocarcinoma cases in TCGA and ONCOSG were used. The best cross-dataset ML performance was reached using transcriptomic data alone and statistically better than those using transcriptomic and clinical data. The best balance accuracy (BA), area under curve (AUC) and accuracy were significantly better in ML algorithms training on TCGA and tested on ONCOSG than those trained on ONCOSG and tested on TCGA ($p<0.05$ for all). Normalization and NDEG greatly improved intra-dataset ML performances in both datasets, but not in cross-dataset testing. Strikingly, modelling transcriptomic data of ONCOSG alone outperformed modelling transcriptomic and clinical data whereas including clinical data in TCGA did not significantly impact ML performance, suggesting limited clinical data value or an overwhelming influence of transcriptomic data in TCGA. Performance gains in intra-dataset testing were more pronounced for ML models trained on ONCOSG than TCGA. Among the six ML models compared, Support vector machine was the most frequent best-performer in both intra-dataset and cross-dataset testing. Therefore, our data show data source, normalization and NDEG are associated with intra-dataset and cross-dataset ML performance in modelling transcriptomic and clinical data.

# Introduction

Intra-dataset (or within dataset) modelling and testing have been commonly used. On the other hand, cross-dataset testing is more robust due to the use of independent dataset, but yet less used. Recently, cross-dataset testing has been used in many fields, including … For example, it can improve understanding of rodent behaviors and interaction [1]. Moreover, the recently developed transfer learning also increased the use of cross-dataset training and testing [2-6]. However, how ML algorithms perform in cross-dataset modelling of transcriptomic and clinical data is largely unknown. Therefore, we first aim to understand the intra-dataset and cross-dataset performance characteristics of ML algorithms in modelling transcriptomic and clinical data.

Lung adenocarcinoma (LUAD) is the predominant subtype of non-small cell lung cancer (NSCLC), accounting for over 40% of all lung cancer cases [7-9]. In recent years, the rapid development of artificial intelligence (AI), machine learning (ML), and deep learning (DL) has provided new approaches for prognostic prediction in LUAD. These models integrate genomic, clinical, and imaging data to assist clinicians in risk assessment and personalized treatment planning [10].

LUAD development and progression are closely related to complex genomic alterations, and genomic data plays a crucial role in prognostic prediction. Gene expression data can be used to predict the prognosis of LUAD patients. MicroRNA and circulating tumor DNA (ctDNA) are important biomarkers for LUAD prognosis [11]. Common mutations, such as those in EGFR, KRAS, and TP53, are frequently observed in LUAD. For example, deep learning models that integrate PET/CT imaging and genomic data can predict EGFR mutation status, thereby guiding targeted therapies [12, 13]Integrating genomic, transcriptomic, and epigenetic data provides a more comprehensive understanding of LUAD's biological characteristics. Deep neural networks built with multi-omics data achieved superior performance, with an AUC of 0.92 in the TCGA dataset [12].

Clinical data includes demographic information, pathology results, treatment plans, and follow-up records. While traditionally applied in statistical models, ML and DL techniques have further unlocked the potential of these data. ML models based on clinical data can identify prognostic factors and predict survival outcomes. For example, Support vector machines and random forests were employed to analyze multicenter clinical data, achieving an AUC of 0.91.[14] Additional research found that smoking history is closely related to prognosis, with smokers experiencing worse outcomes than non-smokers. TNM staging and tumor size are key factors for predicting survival time [15]. Moreover, differences in sex and age also significantly impact prognostic outcomes.

Radiomics and deep learning models have also been widely applied in LUAD survival prediction. Research has shown that CT imaging features, including tumor shape, texture, and edge characteristics, significantly contribute to survival prediction. For example, a predictive model was used to capture survival-relevant imaging patterns [16]. PET/CT fusion imaging offers novel insights into LUAD prognosis. Some research revealed that Metabolic activity in PET scans is closely associated with tumor invasiveness and post-surgical recurrence risk [17]. Similarly, CT- and PET-based deep learning models were utilized to effectively predict LUAD survival by incorporating U-Net and ResNet architectures to improve prediction accuracy [18].

Traditional ML models, such as logistic regression and random forests, have been extensively applied in LUAD survival prediction. For instance, logistic regression models have demonstrated strong performance in five-year survival prediction, with an AUC of 0.88 [11]. DL models have also shown superior performance when handling high-dimensional, nonlinear data. Some CT-based DL models achieved an AUC of 0.93 in survival prediction [18]. Ensemble learning strategies have further enhanced predictive performance. A strategy basing on ensemble learning was proposed to deal with that combined PET and CT data, resulting in improved model performance.[19] Furthermore, LASSO regression, random forests, and XGBoost are widely used for feature selection to identify variables associated with survival outcomes. Different feature selection methods were compared in a research and found that SHAP-based importance measures significantly improved prediction accuracy [20].

Despite remarkable advancements, several challenges persist in LUAD survival prediction, including data heterogeneity, missing values, lack of model interpretability, performance inconsistencies across datasets, and the complexity of molecular subtype prediction. By integrating clinical, imaging, and genomic data, researchers can develop more accurate and personalized prognostic models. With continued efforts in data sharing and algorithm innovation, LUAD survival prediction models are expected to provide more robust clinical support.

# Materials and methods

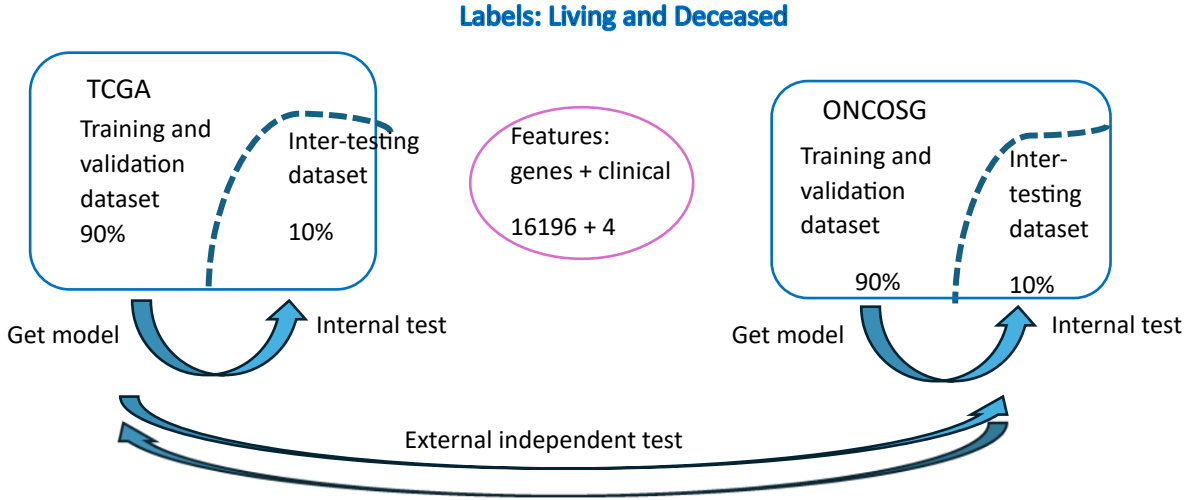

Fig.1 The sample distributions for two datasets

Here, we will propose a cross-dataset analysis framework for predicting cancer patients' overall survival status based on integrating genetic data and clinical data through a series of experiments. The transcriptomic and clinical data of lung adenocarcinoma in the TCGA (the Cancer Genome Atlas) and ONCOSG (Oncology Singapore) were used [21, 22]. The first objective of this study is to investigate the rational selection of stable genes for normalization and differential genes for classification based on the available data. Furthermore, the study will analyze which combinations of normalization and supervised machine learning methods achieve better overall survival prediction. Finally, the study will discuss the role of clinical features in this regard. We trained, validated and test classification models using internal testing within the same dataset and external independent testing in the other dataset.

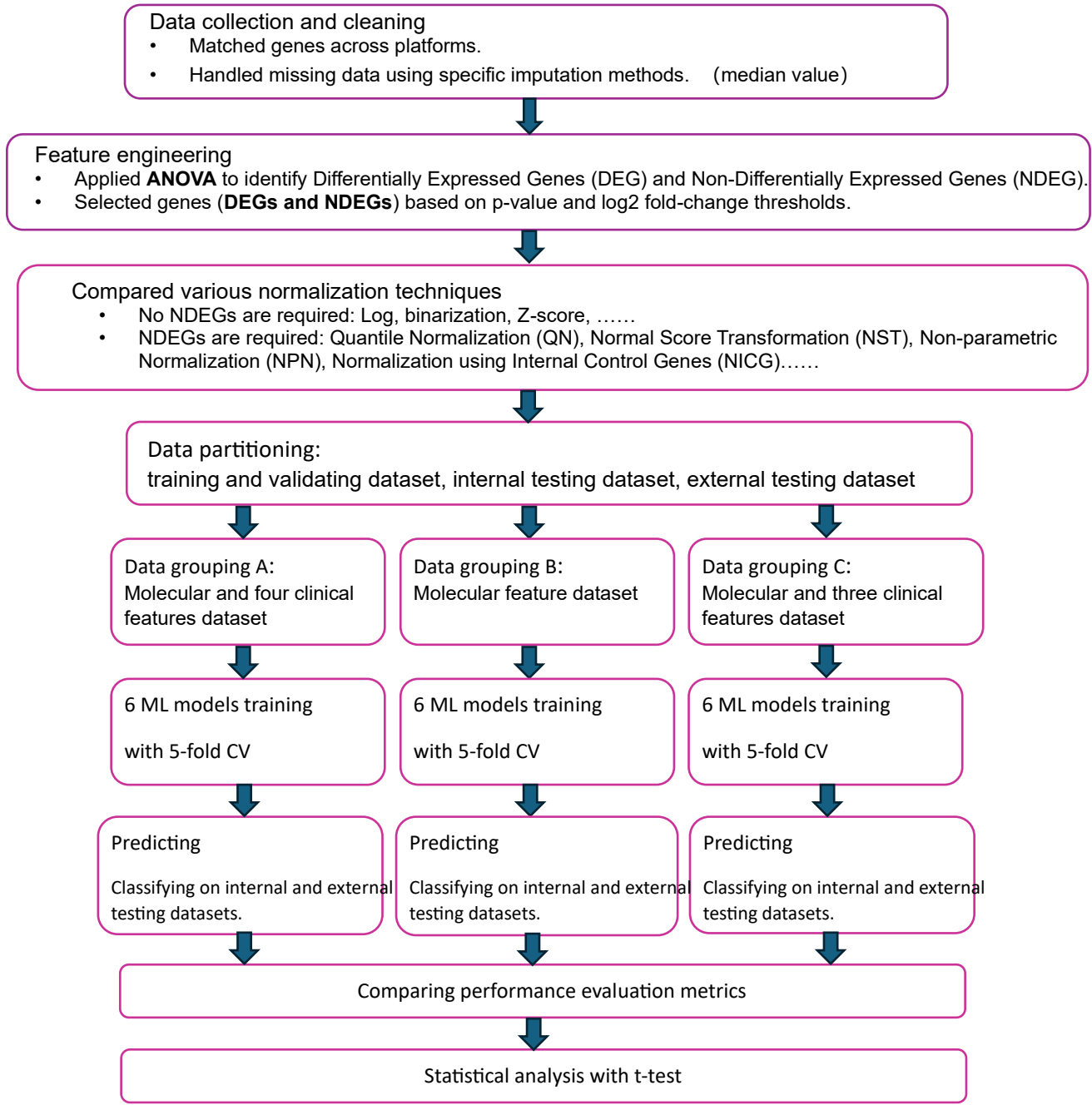

Fig.2 Research framework

Taking this study as an example, we hope to provide researchers with a comprehensive selection strategy for various classification prediction studies based on genetic data. The structure of the paper is description of data used, elaboration of the proposed method, demonstration of results and comparison of models followed by Discussions.

## Dataset Description

As shown in Fig.1, data from TCGA and ONCOSG were used for this study, which include transcriptomic and clinical data. Transcriptomic data is in RNA-seq FPKM format, and further normalized using Z-transformation. The classification goal is binary classification: living and deceased. Both datasets are imbalanced too. For binary classification, sample numbers with living and deceased are 212:74 on TCGA, and 125:42 on ONCSG.

A flowchart for this research is shown in Fig.2. When TCGA data was used for training and validating, we realize the independent internal testing on TCGA data and external testing on ONCOSG data, and vice versa. The specific experimental steps were described before [23]. The entire process was repeated at least five times to obtain a relatively comprehensive model assessment. The experimental steps of each process mainly include: data cleaning, gene selection, normalization, dataset partitioning, classification model training, prediction, classification performance evaluation and Statistical analysis with Student t-test. Python version 3.11.9 64-bit is used for the code implementation.

### Data cleaning

To enable analyses for two datasets, we cleaned the samples by retaining only those with matching labels, keeping shared gene features, and filling missing values with medians. After this preprocessing, the dataset included 16196 gene features and 4 clinical features: age, gender, tumor stage and tumor mutational burden (TMB). These features were chosen because they are shared between two datasets. Some features are numerical, while others are categorical, requiring tailored processing methods. The sample distributions for two datasets are shown in the Fig.1.

### Gene Selection

In this study, the number of samples is significantly smaller than the number of features (16,196 genes), leading to potential multicollinearity and an increased risk of overfitting. The model may fit noise rather than meaningful patterns, increasing computational cost and reducing interpretability. To address these challenges, feature selection or dimensionality reduction is often necessary.

Common feature selection methods include filtering, wrapping, and embedding. Here, we applied one-way ANOVA, a statistical filter method, separately to data from each platform. ANOVA compares between-group variance (differences between category means) and within-group variance (fluctuations within the same category) to determine whether at least one group's mean differs significantly [24]. The F-value, which measures the ratio of these variances, is used to test the null hypothesis that all group means are equal. A high F-value suggests significant differences, making the corresponding genes suitable for classification (differentially expressed genes, DEG), while a low F-value suggests minimal differences, identifying non-differentially expressed genes (NDEG) for normalization [23].

Following statistical principles, F-values are computed from gene expression data and sample category labels and compared to theoretical values in the F-distribution table to determine P-values. By setting different thresholds, gene sets can be defined accordingly. For example, genes with P-values below a selected threshold are designated as DEG for classification, while those above a chosen threshold are designated as NDEG for normalization. We then evaluated how different NDEG and DEG sets influence classification performance by varying these thresholds.

**Data Partitioning**

To fairly evaluate the predictive performance of classification models and their adaptability to cross-platform data, while considering the limited sample size in the dataset, a well-designed data partitioning strategy is required. Here, we employ a bootstrap-based cross-validation method to assess model performance.

In this approach, 90% of randomly selected samples from the TCGA dataset are used for training and five-fold cross-validation, while the remaining 10% serve as an internal test set, then the ONCOSG dataset is used as an external independent test set, and vice versa.

To ensure fairness in model evaluation, the training and test samples remain unchanged throughout a single complete analysis, regardless of variations in gene selection thresholds, normalization methods, or classifier combinations. However, in multiple repeated analyses, the data is randomly re-partitioned each time. Since the data is inherently imbalanced and the sample size is limited, the original class distribution is preserved when splitting the dataset into the training set, validation set, and internal test set. However, during the random partitioning of the external test set, 90% of the samples from the minority class are randomly selected each time, while an equal number of samples from the majority class are randomly chosen to form an independent external test set.

**Normalization**

Normalization is a crucial step in genetic data processing, ensuring comparability across samples. Various normalization methods exist, and the choice depends on the data characteristics and processing goals. Here, we compare several commonly used methods to refine our strategy.

Since the lung cancer transcriptomic data was already Z-transformed (taken as Z_original data), we first examined the effect of classification on both the full raw-dataset and the gene-filtered dataset (taken as Z_raw data). We then evaluated binarization and other four reference gene-based normalization methods: Non-Parametric Normalization (NPN), Quantile Normalization (QN), Quantile Normalization with Z-Score (QN-Z), and Normalization using Internal Control Genes (NICG).[25-28]

- **Z-Score Transformation (Z):** Standardizes data to a mean of 0 and a standard deviation of 1, making features comparable.

- **Non-Parametric Normalization (NPN):** A robust method that ranks data and maps it to a reference distribution without assuming a specific data distribution.

- **Quantile Normalization (QN):** Aligns expression distributions across samples to reduce technical variability. When applied using reference genes (RQN), it improves comparability by adjusting based on stably expressed genes.

- **Normalization using Internal Control Genes (NICG):** Uses housekeeping genes as a normalization factor to correct technical variations.

- **QN-Z:** Applies QN followed by Z-transformation for further standardization.

Each normalization method was applied consistently to both training and test datasets. We assessed their impact on classification performance by comparing results against direct classification on the original data.

**Machine learning Models**

We trained six common machine learning classifiers on different training sets: Multilayer Perceptron (MLP), Extreme Gradient Boosting (XGBoost), Logistic Regression (LR), Least Absolute Shrinkage and Selection Operator (LASSO), Support Vector Machine (SVM), and Random Forest (RF). These models, widely used in practice, differ in characteristics, making them suitable for analyzing dataset-model interactions.

- **LR[29]**: A linear model that balances interpretability and efficiency, suitable for datasets with many features.

- **LASSO[30]**: A regression model with L1 regularization, which performs feature selection by shrinking some coefficients to zero, reducing overfitting and enhancing interpretability. It is particularly effective for high-dimensional datasets.

- **MLP[31, 32]**: A basic deep learning model with hidden layers, capable of capturing complex nonlinear relationships in classification tasks.

- **RF[33]**: An ensemble of decision trees that improves robustness by introducing random feature selection, effective for handling nonlinear relationships and outliers.

- **XGBoost[34, 35]**: A gradient boosting model that optimizes regularization and prevents overfitting, excelling in structured and sparse data.

- **SVM[36]**: A supervised algorithm that finds the optimal hyperplane for classification. It handles high-dimensional and nonlinear problems well using kernel tricks.

Considering the imbalance in the dataset, class weights were applied in the XGB and SVM models, referred to as XGB_W and SVM_W, respectively. Given the differences among models in terms of data dimensionality, feature interactions, and sparsity, we conducted a comprehensive evaluation of each model's adaptability to the dataset using the same data.

**Classification Performance Evaluation**

Due to the multi-class and unbalanced nature of the data in this study, a combination of balanced accuracy and the Kappa statistic, in addition to F1, AUC, sensitivity, and specificity, was primarily used to evaluate classification performance based on the internal and external testing set.[37, 38]

Balanced accuracy is a metric that accounts for class imbalance and represents the average accuracy for each class. In the case of an unbalanced dataset, the overall accuracy may be high, despite the fact that the predictions for a few classes may be inaccurate. Balanced accuracy provides a fairer assessment of the model's performance across all classes. It is calculated as:

$$\text{Balanced Accuracy} = (1/n) \sum_{i=1}^{n} \left(\frac{\text{True Positives i}}{\text{Total Class i}}\right), \quad (1)$$

where n is the number of classes.

# Results

## ML model performance using transcriptomic and all clinical data

We repeated the processing flow in Fig.2 five times to obtain average performance metrics. Here, we visualize the average balanced accuracies of these optimal models to facilitate comparison and analysis.

## Results of internal testing

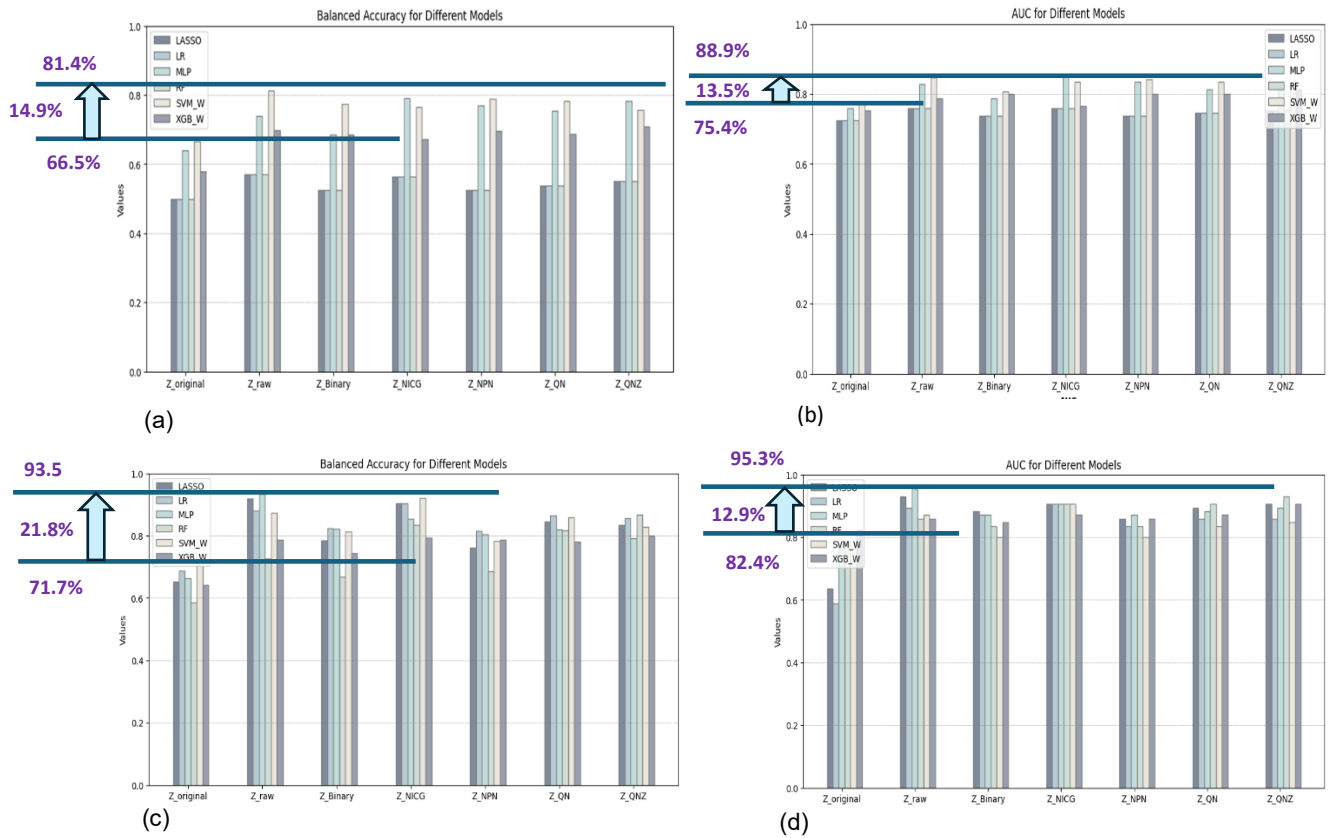

Fig.3 The internal testing results on transcriptomic and four clinical features (Data grouping A)

a) Balanced accuracy on TCGA dataset; b) AUC on TCGA dataset; c) Balanced accuracy on ONCOSG dataset; d) AUC on ONCOSG dataset.

All expression values corresponding to the 16196 genes and 4 clinical features shared by the two datasets are directly used for the analysis to observe the performance of the six classifiers in the original data (Z-transformed data), raw data (Z-transformed data further processed by gene selection strategy) or the Z-transformed data further processed by different reference gene-based normalization methods.

The results of internal testing conducted on the two datasets are shown in Fig.3. The trends observed across different combinations of normalization methods and classification models exhibit similarities, indicating that the relationship between the information conveyed by the data and the classification objective shares certain consistencies. For the binary classification task performed on the TCGA dataset, the balanced accuracy increased from 66.5% in the Z-original data to 81.4%, an improvement of nearly 15 percentage points. The best results were achieved on the Z-raw data. AUC also improved by 13.5 percentage points. Since balanced accuracy takes the average recall across all classes, it is more suitable for evaluating model performance on imbalanced datasets. A similar trend was observed in the internal testing results on the ONCOSG dataset. However, the ONCOSG results were generally superior to those of TCGA: the balanced accuracy improved by 21.8 percentage points, and AUC increased by 12.9 percentage points.

We found that all classification models showed significant performance improvements on Z-raw data compared to the Z-original data, indicating that our DEG-NDEG gene selection strategy played a crucial role in enhancing classification performance. However, further implementing additional normalization methods on Z-raw data did not yield further significant improvements and, in some cases, even led to performance degradation. We believe this is mainly because lung cancer transcriptomic data had already undergone Z-transformation, which altered its original distribution, making the impact of additional normalization strategies very limited.

When comparing the performance of different classification models across the two datasets, we also found that MLP and SVM consistently performed relatively well in all scenarios. This is because MLP and SVM are better suited for handling nonlinear relationships. The SVM model optimizes the maximum-margin hyperplane, making it inherently robust against individual noisy features. Meanwhile, MLP, through nonlinear activation functions, learns complex feature relationships and can automatically adjust weights, making it less sensitive to noisy features. Our classification task particularly benefits from the flexibility of models like MLP and SVM, which excel in exploring complex relationships.

## Results of the external independent testing

The external independent testing results between two datasets are shown in Fig. 4.

The differences between datasets from different platforms, caused by various factors, have long been a challenging issue for machine learning models to overcome. As a result, when using a model trained on the TCGA data to predict ONCOSG samples, the increased difficulty led to a performance decline as expected: the balanced accuracy improved by 10.6 percentage points, and AUC increased by 7.0 percentage points. However, when using a model trained on ONCOSG data to predict TCGA samples, the results were disastrous—the balanced accuracy and AUC showed almost no improvement and remained only slightly above 0.5. This indicates that for this cross-platform classification task, the effectiveness of all models was extremely limited.

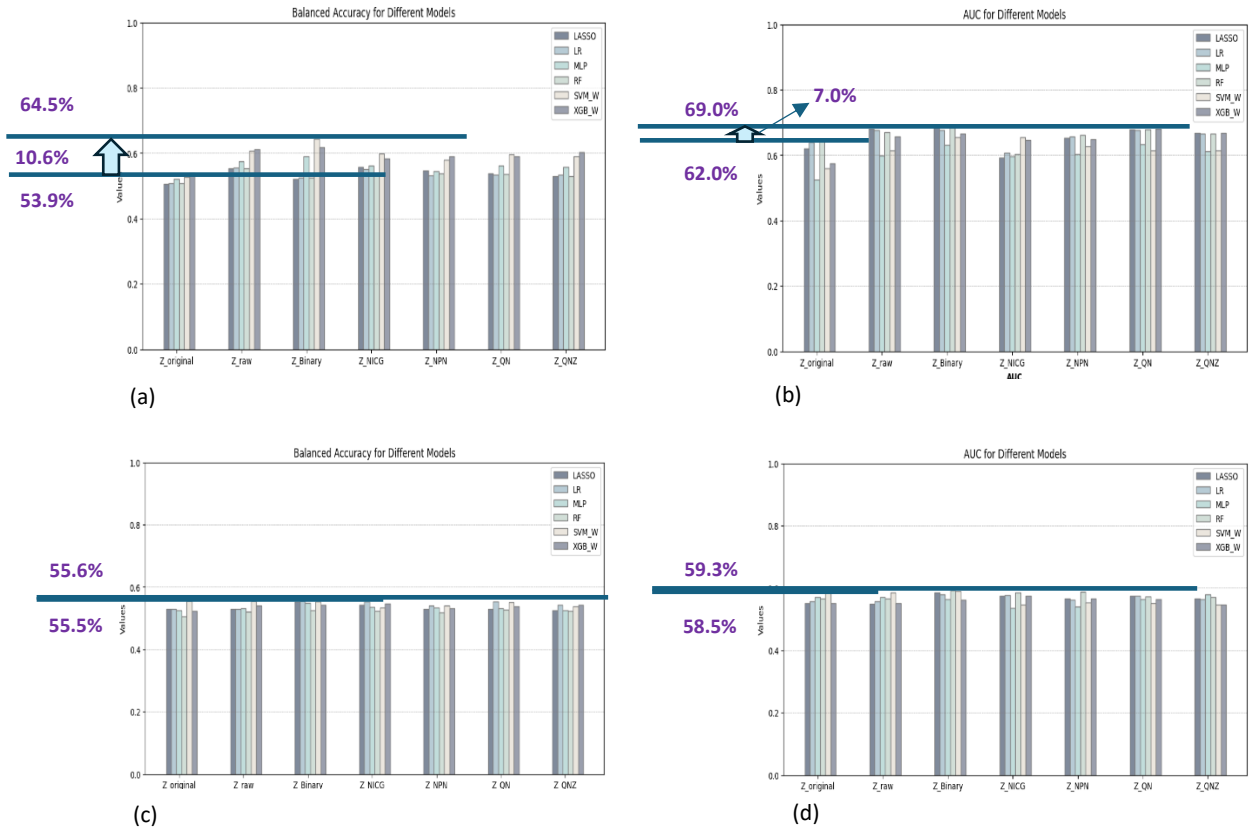

Fig.4 The external independent testing results on transcriptomic and four clinical features (Data grouping A)

a) Balanced accuracy predicted on ONCOSG data with model trained on TCGA data; b) AUC predicted on ONCOSG data with model trained on TCGA data; c) balanced accuracy predicted on TCGA data with model trained on ONCOSG data; d) AUC predicted on TCGA data with model trained on ONCOSG data.

## ML model performance using transcriptomic and three clinical features

In this study, we selected four clinical features: age, gender, tumor stage, and TMB. These features were chosen because they are the common clinical features shared between the two datasets we used. However, are these features crucial for our classification task?

To explore this, we tried to remove tumor stage, which seems to be the feature most closely related to prognosis, while keeping all other conditions unchanged. We then looked at the internal and external binary classification results of the corresponding models shown in Figure 5 and Figure 6, respectively, and analyzed the relationship between these clinical features and the classification goal.

## Results of internal testing

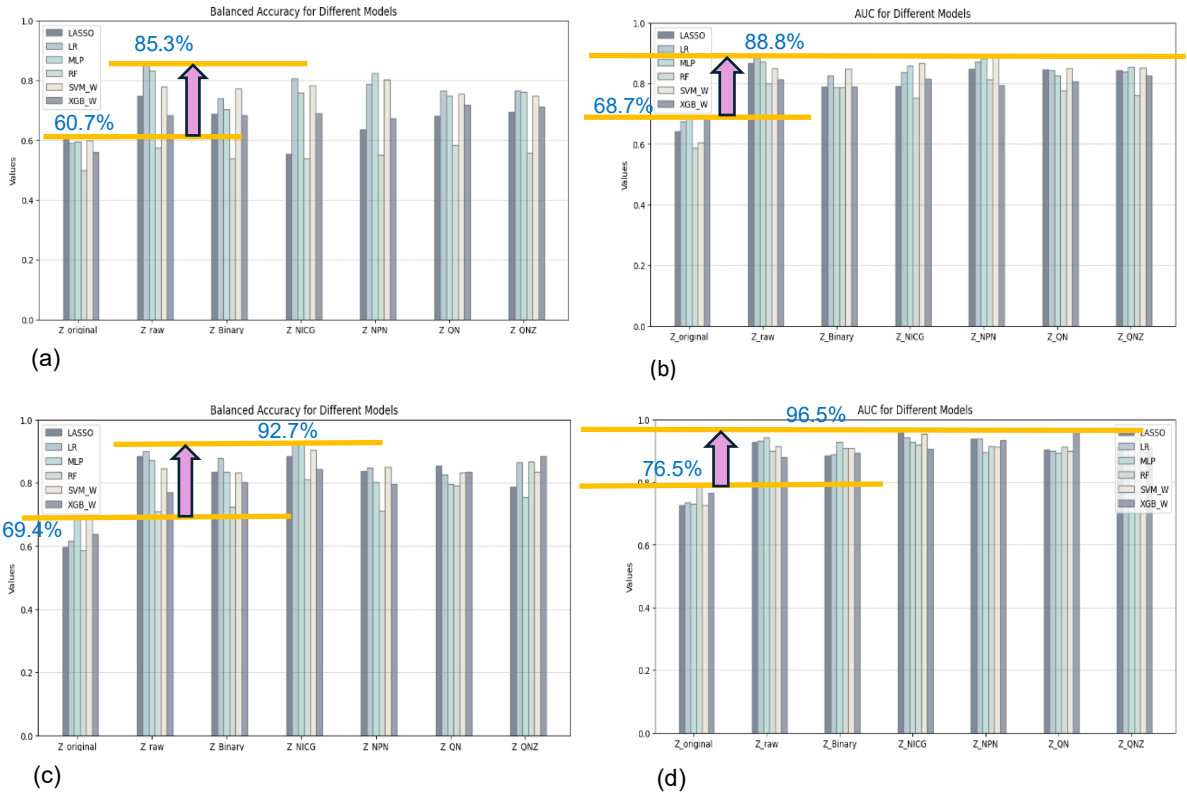

Fig.5 The internal testing results transcriptomic and three clinical data (Data grouping C)

a) Balanced accuracy on TCGA dataset; b) AUC on TCGA dataset; c) Balanced accuracy on ONCOSG dataset; d) AUC on ONCOSG dataset.

Figure 5 presents the internal testing results obtained on datasets from both platforms, demonstrating that the removal of the tumor stage feature did not lead to a dramatic decline in model performance. On the TCGA dataset, the performance of the LR model and LASSO model has been greatly improved. The LR model achieves the best results on the Z-raw data: the balanced accuracy still improved from 60.7% on the Z-original data to 85.3%, while the AUC increased from 69.4% to 93.5%. Similarly, the performance on the ONCOSG dataset remained superior to that on the TCGA dataset, and the best effect is achieved on Z-NICG, with the balanced accuracy increasing from 69.4% on the z-original data to 92.7% and the AUC rising from 82.4% to 100%.

## Results of external independent testing

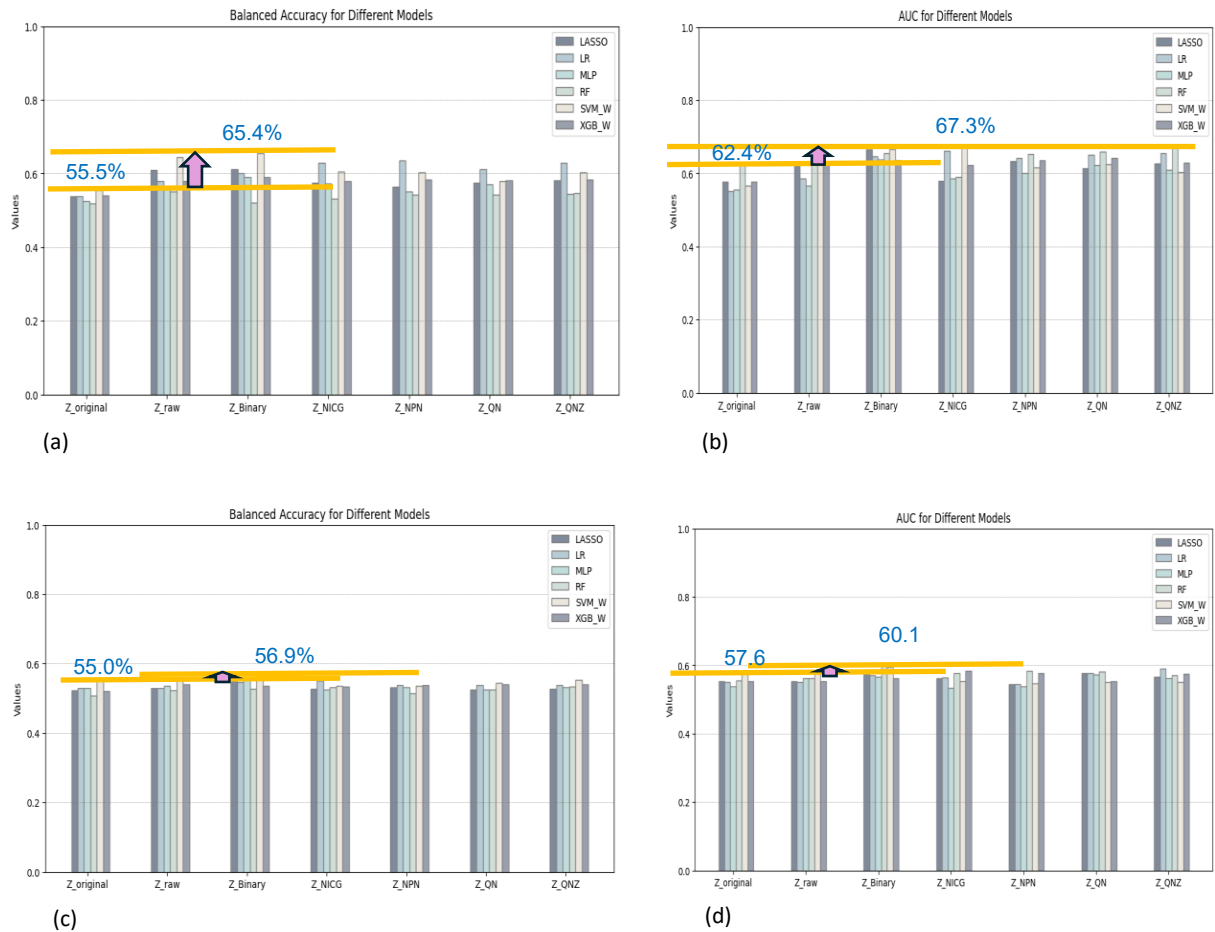

Fig.6 The external independent testing results on transcriptomic and three clinical features (Data grouping C)

a) Balanced accuracy predicted on ONCOSG data with model trained on TCGA data; b) AUC predicted on ONCOSG data with model trained on TCGA data; c) balanced accuracy predicted on TCGA data with model trained on ONCOSG data; d) AUC predicted on TCGA data with model trained on ONCOSG data.

After removing the tumor stage feature, the results observed in cross-platform external independent testing remained consistent: when the model trained on the TCGA dataset was applied to ONCOSG, the balanced accuracy still improved by nearly 10 percentage points, while the AUC increased by approximately 7 percentage points. However, when the model trained on the ONCOSG dataset was applied to TCGA data, although the AUC improved by nearly 10 percentage points, the balanced accuracy showed almost no improvement.

## ML model performance using transcriptomic features alone

Since the model's performance showed little change after removing the tumor stage feature, we were surprised and encouraged to further remove all clinical features to observe the resulting changes. The internal and external testing results obtained using only transcriptomic data are presented in Figures 7 and 8, respectively.

**Results of internal testing**

Using only transcriptomic data, the internal testing results on datasets from both platforms remain satisfactory. On the TCGA dataset, the balanced accuracy increased from 64.9% in the z-original data to 84.8%, while the AUC improved from 68.7% to 88.8%. On the ONCOSG dataset, the balanced accuracy increased from 74.2% in the z-original data to 97.7%, and the AUC improved from 76.5% to 96.5%. Notably, except for the Z-Raw data, the performance of the LR, MLP, and SVM_W models showed significant improvements on the Z-NPN and Z-NICG datasets.

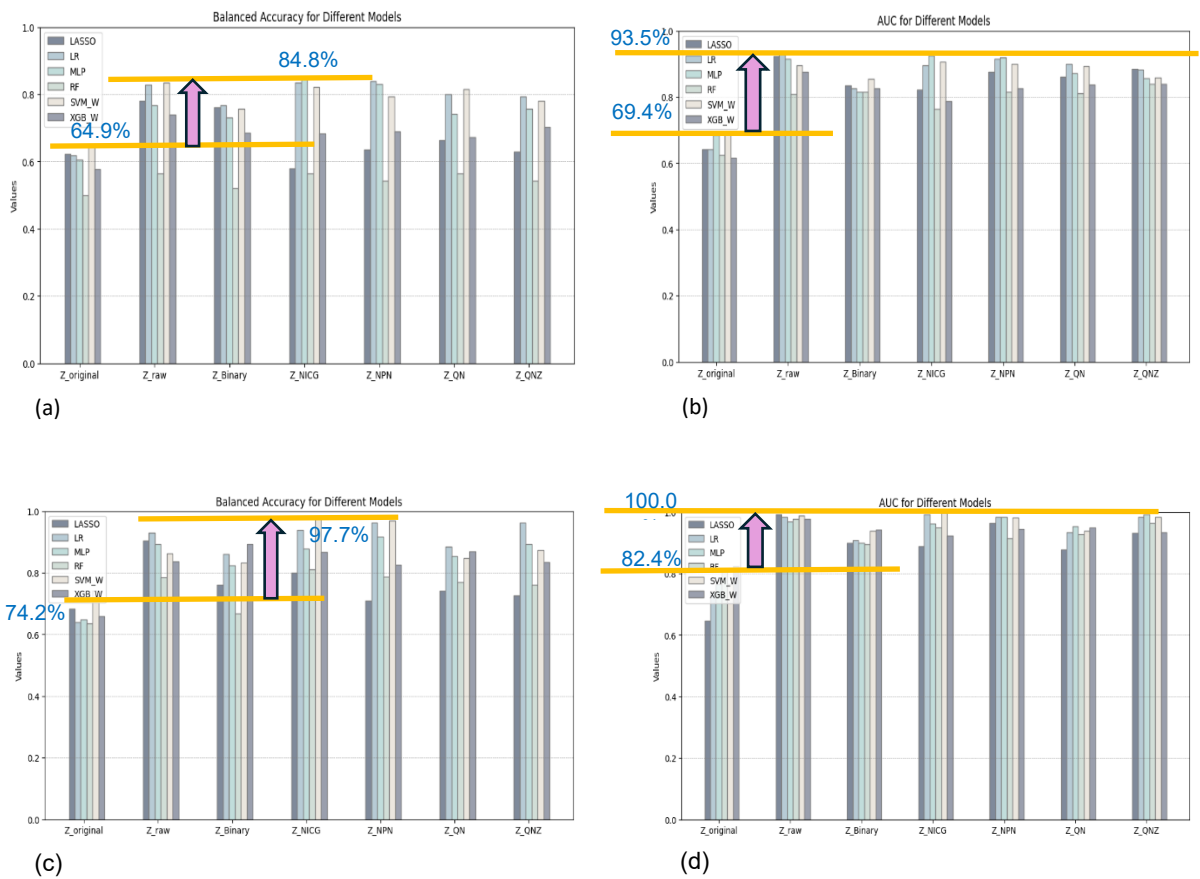

Fig.7 The internal testing results on only transcriptomic data (Data grouping B)

a) Balanced accuracy on TCGA dataset; b) AUC on TCGA dataset; c) Balanced accuracy on ONCOSG dataset; d) AUC on ONCOSG dataset.

**Results of external independent testing**

After removing all clinical features, the cross-platform external independent testing results showed that when the model trained on the TCGA dataset was applied to ONCOSG, the balanced accuracy improved by approximately 7 percentage points, and the AUC increased by about 5 percentage points. Similarly, when the model trained on the ONCOSG dataset was applied to TCGA, both the balanced accuracy and AUC showed a slight improvement of around 3 percentage points.

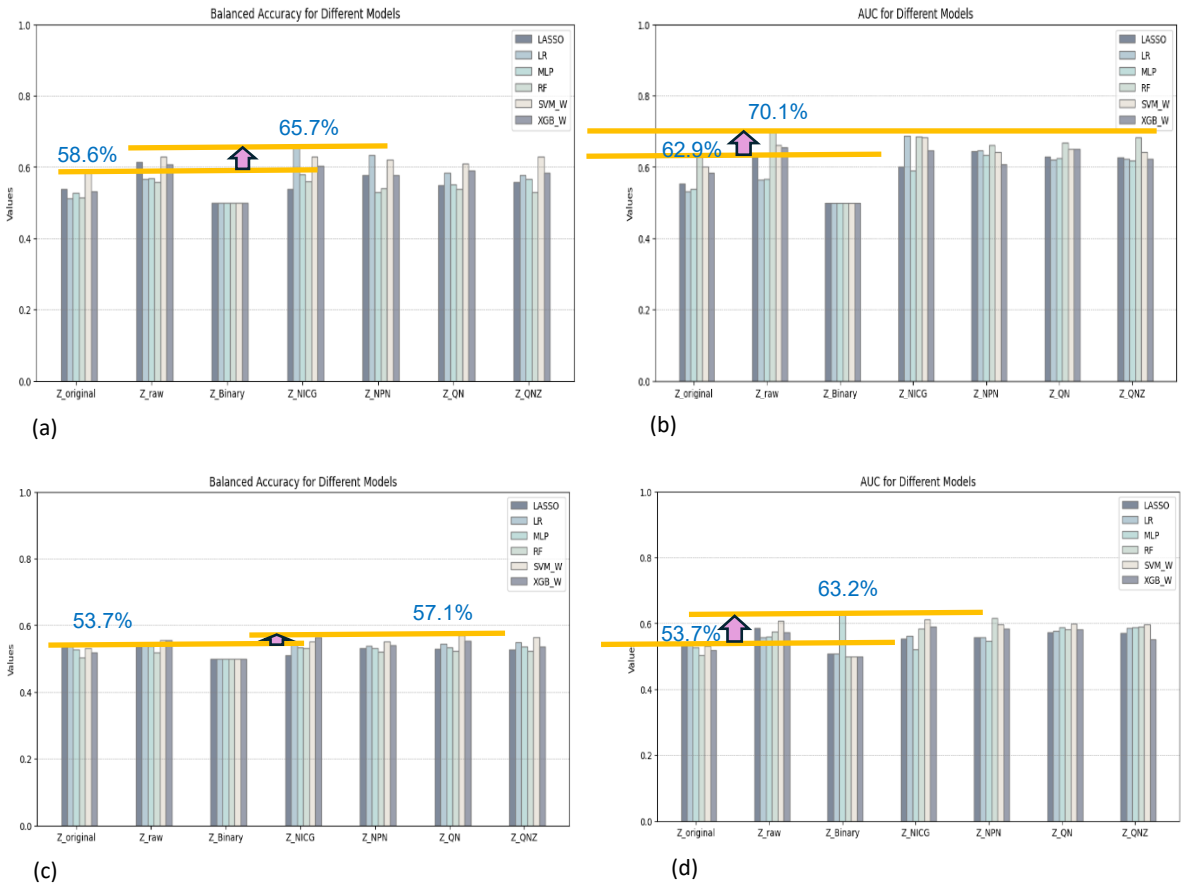

Fig.8 The external independent testing results on only transcriptomic data (Data grouping B)

a) Balanced accuracy predicted on ONCOSG data with model trained on TCGA data; b) AUC predicted on ONCOSG data with model trained on TCGA data; c) Balanced accuracy predicted on TCGA data with model trained on ONCOSG data; d) AUC predicted on TCGA data with model trained on ONCOSG data.

## Statistical Analysis

Since we computed the model performance evaluation metrics by conducting classification predictions on multiple randomly selected sample combinations from internal or external test sets and subsequently calculated the mean as the final evaluation result, we conducted a t-test on the multiple evaluation results to assess whether the classification results obtained under the three modes exhibit statistical significance. The average balanced accuracy and the corresponding p-values obtained from the t-test for each model are shown in the supplementary table 1-16 respectively.

The above results suggest that the inclusion of clinical features does not have a particularly significant impact on model performance. However, models trained on the TCGA dataset and the ONCOSG dataset exhibit different performances in external validation. Since we calculate model performance evaluation metrics by performing classification predictions on multiple randomly selected sample combinations from the internal or external test sets and then compute the mean as the final evaluation result, we conducted t-tests on the best-performing models under three different conditions to further determine whether the observed differences in average performance are due to random fluctuations or have statistical significance.

For narrative convenience, we refer to the model based on genetic features and four clinical feature data as Data grouping A, the model using only genetic feature data as Data grouping B, and the one based on genetic features and three clinical feature data as Data grouping C.

First, we compared best internal testing performances of models trained on the TCGA dataset with those trained on the ONCOSG dataset under the three conditions mentioned above. The results, shown in Table 1, indicate that when only transcriptomic data is used, the performance differences between the two datasets using the same method are significant and statistically meaningful. Moreover, the statistical significance of this difference is even more pronounced in cross-platform external testing. Models trained on the TCGA dataset exhibit significantly better predictive performance on the ONCOSG dataset compared to models trained on the ONCOSG dataset when tested on the TCGA dataset. This discrepancy may stem from the fact that the ONCOSG dataset primarily consists of samples from Asian populations. We aim to further investigate in future studies whether this outcome is influenced by racial differences.

Table 1. Compare the best internal testing performance of models trained on TCGA data versus those trained on ONCOSG data.

|  |  | Data grouping A | | | Data grouping B | | | Data grouping C | | |
|---|---|---|---|---|---|---|---|---|---|---|
|  |  | TCGA as training set | ONCOSG as training set | P-value | TCGA as training set | ONCOSG as training set | P-value | TCGA as training set | ONCOSG as training set | P-value |
| Internal testing | Balanced accuracy | 0.814± 0.010 | 0.935± 0.004 | 0.057 | 0.848± 0.001 | 0.977± 0.000 * | 0.0004 | 0.853± 0.011 | 0.927± 0.003 | 0.211 |
|  | AUC | 0.888± 0.023 | 0.953± 0.002 | 0.403 | 0.925± 0.019 | 1.000± 0.000 * | 0.291 | 0.885± 0.008 | 0.912± 0.010 | 0.665 |
|  | Accuracy | 0.821± 0.006 | 0.977± 0.001 | 0.008 | 0.890± 0.001 | 0.965± 0.001 | 0.006 | 0.910± 0.005 | 0.941± 0.002 | 0.436 |
|  | DEG | 78 | 534 |  | 1430 | 2382 |  | 996 | 2070 |  |

|  | | | | | | | | | |
|---|---|---|---|---|---|---|---|---|---|
| | NDEG | 62 | 230 | | 120 | 65 | | 120 | 65 |
| | Normalization method | Z-Raw | Z-Raw | | Z-NPN | Z-NICG | | Z-Raw | Z-NICG |
| | Classification model | SVM_W | MLP | | MLP | SVM_W | | LR | MLP |
| External independent testing | Balanced accuracy | 0.645±0.003 | 0.556±0.000 * | 0.022 | 0.657±0.001 | 0.571±0.000 * | 0.004 | 0.654±0.001 | 0.569±0.000 * | 0.004 |
| | AUC | 0.654±0.002 | 0.579±0.000 * | 0.020 | 0.687±0.001 | 0.599±0.000 * | 0.003 | 0.665±0.002 | 0.595±0.000 * | 0.025 |
| | Accuracy | 0.645±0.003 | 0.556±0.000 * | 0.022 | 0.657±0.001 | 0.571±0.000 * | 0.004 | 0.654±0.001 | 0.569±0.000 * | 0.004 |
| | DEG | 161 | 617 | | 176 | 2382 | | 816 | 2960 |
| | NDEG | 120 | 1729 | | 120 | 230 | | 120 | 562 |
| | Normalization method | Z-Binary | Z-Binary | | Z-NICG | Z-QN | | Z-Binary | Z-Binary |
| | Classification model | SVM_W | LR | | LR | SVM_W | | SVM_W | SVM_W |

All data are shown in mean ± standard deviation. **Data grouping: A**, transcriptomic data plus 4 clinical features; **B**, transcriptomic data alone; **C**, transcriptomic data plus 3 clinical features. All experiments were repeated 15 times. **\*** The SD was less than 0.00001. **DEG**, Differentially expressed genes; **NDEG**, non-differentially expressed genes; **Z-Raw**, data with gene selection strategies applied based on the original data (Z-transformed data); **Z-Binary**, Data obtained after binarization based on Z-Raw data; **Z-QN**, data with Quantile Normalization applied on the Z-Raw data; **Z-NICG**, data with Normalization using Internal Control Genes applied on the Z-Raw data; **Z-NPN**, data with Non-Parametric Normalization applied on the Z-Raw data; **SVM_W**, class weights were applied in the Support Vector Machine model; **LR**, Logistic Regression; **MLP**, Multilayer Perceptron.

In addition, we also compared the best performance of ML in internal testing and that in external testing obtained on data groupings A, B, and C (Supplementary Tables 17–18). Interestingly, the t-test results indicate that no cases exhibit statistically significant differences, while the prediction performance of models trained on ONCOSG data and applied to TCGA data shows significant differences in the three conditions.

## Discussion

This secondary study focused on the intra-dataset and cross-dataset performances of ML using transcriptomic data with or without clinical data. The best cross-dataset ML performance on predicting overall survival of LUAD patients was reached using transcriptomic data alone and was statistically better than those using transcriptomic and clinical data. The best BA, AUC and accuracy were 0.657±0.001 (mean ± standard deviation), 0.687±0.001, and 0.657±0.001 for training on TCGA, 0.571±<0.0001, 0.599± <0.0001 and 0.571±<0.0001 for training on ONCOSG, respectively. These best performance metrics were statistically different between those trained on TCGA and those trained on ONCOSG, with the difference of 0.086 in BA, 0.088 in AUC and 0.086 in accuracy. We also found that normalization and selecting NDEG greatly improved ML performance in internal (intra-dataset) testing, but not so in external (cross-dataset) testing.

The differences in ML performance observed in external (cross-dataset) testing are also reflected in the fact that models trained on ONCOSG performed worse than those trained on TCGA. This discrepancy may stem from the fact that the ONCOSG dataset primarily consists of samples from Asian populations [21], whereas the TCGA dataset includes a racially more diverse population [22]. The limited representativeness of the ONCOSG dataset may directly impact the cross-dataset ML performance due to its more homogenous patient population and less diverse data source than the TCGA.

We here show that cross-dataset testing on different datasets may change various ML performance metrics in modelling transcriptomic data with or without clinical data. There are to our knowledge no published cross-dataset studies on this subject, while several cross-dataset ML studies in other fields support our findings [2, 4, 39]. The reported differences in accuracy were 0.018 (76.82 to 71.05%), 0.07 (0.97 to 0.91) and 0. 35(0.91 to 0.56), respectively [2, 4, 40]. The differences in AUC ranged from 0.06 (0.94 to 0.88) to 0.27 (0.99 to 0.72) [41, 42]. However, our study is novel in that we comprehensively examined various performance metrics, such as balanced accuracy, AUC and balance. Others have only compared accuracy or AUC and some only reported one of them. Moreover, we repeated our experiments 15 times and reported statistics of the metrics. Furthermore, we conducted inferential analyses to rigorously compare the ML performance. Finally, most studies reported accuracy or AUC as the ML performance metric. However, we comprehensively examined all the three ML performance metrics (BA, AUC and accuracy) and found some differences among these metrics. To address the issue of performance difference, standardization of the datasets may be a good start. Indeed, it has been proposed and realized for smartphone-based human activity recognition [43].

One key strength of this study is the use of six ML algorithms, while few ML classification studies that used intra-dataset and cross-dataset testing on transcriptomic and clinical data with 6 ML algorithms. Indeed, most published cross-dataset studies only used one or two commonly used ML algorithms [4]. Among the six ML algorithms, SVM was in this study the most frequent top performer of the ML in intra-dataset (internal) and cross-dataset (external) testing (Table 1). This is supported by a prior study, in which normalization seems to improve performance of SVM, but not artificial neural network [44]. SVM also outperformed LR in predicting antidepressant responses to non-invasive brain stimulation with electroencephalogram data [45]. Interestingly, RF appears to reach better accuracy than SVM and LR in the same study [45].

This study has several noteworthy limitations. First, the ML performance on cross-dataset testing was overall moderate (BA, AUC and accuracy were 0.657, 0.687, 0.657, respectively), although normalization and selecting NDEG have improved the performance in selected models. Second, the available clinical features that were overlapped in both TCGA and ONCOSG are limited to four. Despite that transcriptomic data alone may be sufficient to reach the highest ML performance, it will be interesting and perhaps necessary to conduct similar studies on the datasets with more clinical features. Interestingly, we prior work on the intra-dataset testing also showed the clinical features in TCGA may not be required to reach the best ML performance on predicting deaths in LUAD patients [8]. Third, we limited the test sample to a balanced composition so that the ML performance can be more robustly compared. However, this may not mimic the real-world scenarios and tends to produce better ML performance. Finally, the use of normalization and selecting NDEG may overfit intra-dataset data and underperform in cross-dataset testing. It is not very concerning because these approaches did not appear to greatly improve ML performance in cross-dataset testing. Future works are required to examine our findings in separate datasets and disease models.

There are in our view two possible explanations for the lack of strong influence of clinical features on the ML performance, especially in cross-dataset testing. First, the expression levels of certain genes may be strongly correlated with clinical features such as age, tumor stage, and gender, resulting in no additional information gain from these clinical features in the model. This could even lead to redundant information, thereby increasing noise in the model. Second, from a biological perspective, gene features are usually able to capture transcriptomic-level signals that directly reflect disease progression or prognosis, while clinical features merely represent macroscopic phenomena. For instance, clinical features such as age and gender are merely external manifestations of these transcriptomic changes. If the gene data already captures deep biological signals, the contribution of clinical features may become negligible.

# Conclusion and future works

This work systematically evaluated the association of normalization, selecting NDEG and data source with machine learning performance in intra-dataset or cross-dataset modelling of transcriptomic and clinical data. Our data show that normalization and selecting NDEG can improve intra-dataset ML modelling of these data, but not cross-dataset ML performance. There was also significant difference in ML performance in different cross-dataset testing, suggesting it is important or necessary to cross train ML algorithms in each dataset and test on one of the other(s). Future works should focus on how to best understand and reduce the difference in the cross-dataset testing when using different training datasets.

# Acknowledgments

# Funding

This work was supported by the National Cancer Institute, National Institutes of Health (grant number R37CA277812 to LZ). The funder of the study had no role in study design, data collection, data analysis, data interpretation, or writing of the report. The corresponding author had full access to all the data in the study and had final responsibility for the decision to submit for publication.

# Author Contributions Statement

Study conceptualization and design, ensuring the data access, accuracy and integrity (LZ), and manuscript writing (FD and LZ). Both authors contributed to the writing or revision of the review article and approved the final publication version.

# Conflicts of Interest

The authors declare no other conflict of interests.

# Data Availability Statement

The data sets used and/or analyzed of this study are available on the cBioPortal website (https://www.cbioportal.org/). The program coding is available from the corresponding authors on reasonable request.

# Compliance with ethical standards

This exempt study using publicly available de-identified data did not require an IRB review.

[32] R.A. Dunne, A statistical approach to neural networks for pattern recognition, John Wiley & Sons2007.
[33] A. Parmar, R. Katariya, V. Patel, A review on random forest: An ensemble classifier, International conference on intelligent data communication technologies and internet of things (ICICI) 2018, Springer, 2019, pp. 758-763.
[34] B. Ma, F. Meng, G. Yan, H. Yan, B. Chai, F. Song, Diagnostic classification of cancers using extreme gradient boosting algorithm and multi-omics data, Comput Biol Med, 121 (2020) 103761.
[35] R.P. Sheridan, W.M. Wang, A. Liaw, J. Ma, E.M. Gifford, Extreme gradient boosting as a method for quantitative structure–activity relationships, Journal of chemical information and modeling, 56 (2016) 2353-2360.
[36] Y.-M. Huang, S.-X. Du, Weighted support vector machine for classification with uneven training class sizes,  2005 international conference on machine learning and cybernetics, IEEE, 2005, pp. 4365-4369.
[37] Ž. Vujović, Classification model evaluation metrics, International Journal of Advanced Computer Science and Applications, 12 (2021) 599-606.
[38] S. Raschka, Model evaluation, model selection, and algorithm selection in machine learning. arXiv 2018, arXiv preprint arXiv:1811.12808, (2021).
[39] A.C. Yu, B. Mohajer, J. Eng, External Validation of Deep Learning Algorithms for Radiologic Diagnosis: A Systematic Review, Radiol Artif Intell, 4 (2022) e210064.
[40] G. Ben Or, I. Veksler-Lublinsky, Comprehensive machine-learning-based analysis of microRNA-target interactions reveals variable transferability of interaction rules across species, BMC Bioinformatics, 22 (2021) 264.
[41] T. Mohammadzadeh-Vardin, A. Ghareyazi, A. Gharizadeh, K. Abbasi, H.R. Rabiee, DeepDRA: Drug repurposing using multi-omics data integration with autoencoders, PLoS One, 19 (2024) e0307649.
[42] L. Giancardo, F. Meriaudeau, T.P. Karnowski, Y. Li, S. Garg, K.W. Tobin, Jr., E. Chaum, Exudate-based diabetic macular edema detection in fundus images using publicly available datasets, Med Image Anal, 16 (2012) 216-226.
[43] O. Napoli, D. Duarte, P. Alves, D.H.P. Soto, H.E. de Oliveira, A. Rocha, L. Boccato, E. Borin, A benchmark for domain adaptation and generalization in smartphone-based human activity recognition, Sci Data, 11 (2024) 1192.
[44] B. Koo, J. Kim, Y. Nam, Y. Kim, The Performance of Post-Fall Detection Using the Cross-Dataset: Feature Vectors, Classifiers and Processing Conditions, Sensors (Basel), 21 (2021).
[45] C.T. Li, C.S. Chen, C.M. Cheng, C.P. Chen, J.P. Chen, M.H. Chen, Y.M. Bai, S.J. Tsai, Prediction of antidepressant responses to non-invasive brain stimulation using frontal electroencephalogram signals: Cross-dataset comparisons and validation, J Affect Disord, 343 (2023) 86-95.

# Supplementary Tables

Supplementary Table 1. Performance metrics of internal testing obtained by models trained on TCGA data with molecular and 4 clinical features (age, gender TMB and tumor stage) (Data grouping A)

| Balanced Accuracy | LASSO | LR | MLP | RF | SVM_W | XGB_W |
|---|---|---|---|---|---|---|
| Z_Original | 0.5 | 0.5 | 0.64 | 0.5 | 0.665 | 0.58 |
| Z_Raw | 0.57 | 0.57 | 0.74 | 0.57 | **0.814** | 0.698 |
| Z_Binary | 0.525 | 0.525 | 0.685 | 0.525 | 0.774 | 0.685 |
| Z_NICG | 0.563 | 0.563 | **0.792** | 0.563 | 0.765 | 0.672 |
| Z_NPN | 0.525 | 0.525 | 0.77 | 0.525 | 0.79 | 0.696 |
| Z_QN | 0.538 | 0.538 | 0.755 | 0.538 | 0.782 | 0.687 |
| Z_QNZ | 0.55 | 0.55 | **0.783** | 0.55 | 0.757 | 0.709 |

| AUC | LASSO | LR | MLP | RF | SVM_W | XGB_W |
|---|---|---|---|---|---|---|
| Z_Original | 0.607 | 0.607 | 0.754 | 0.607 | 0.645 | 0.656 |
| Z_Raw | 0.776 | 0.776 | 0.857 | 0.776 | **0.888** | 0.806 |
| Z_Binary | 0.785 | 0.785 | 0.796 | 0.785 | 0.845 | 0.779 |
| Z_NICG | 0.742 | 0.742 | 0.838 | 0.742 | **0.889** | 0.774 |
| Z_NPN | 0.754 | 0.754 | 0.852 | 0.754 | **0.871** | 0.783 |
| Z_QN | 0.756 | 0.756 | 0.842 | 0.756 | 0.838 | 0.788 |
| Z_QNZ | 0.787 | 0.787 | 0.836 | 0.787 | 0.817 | 0.812 |

Supplementary Table 2. Performance metrics of internal testing obtained by models trained on TCGA data with molecular features (Data grouping B)

| Balanced Accuracy | LASSO | LR | MLP | RF | SVM_W | XGB_W |
|---|---|---|---|---|---|---|
| Z_Original | 0.622 | 0.619 | 0.606 | 0.5 | 0.649 | 0.577 |
| Z_Raw | 0.781 | 0.829 | 0.768 | 0.563 | **0.834** | 0.739 |
| Z_Binary | 0.76 | 0.767 | 0.731 | 0.521 | 0.756 | 0.685 |
| Z_NICG | 0.58 | **0.835*** | 0.848 | 0.563 | 0.821 | 0.682 |
| Z_NPN | 0.635 | **0.838*** | 0.83 | 0.542 | 0.794 | 0.689 |
| Z_QN | 0.664 | 0.801 | 0.741 | 0.563 | 0.816 | 0.672* |
| Z_QNZ | 0.63 | 0.794* | 0.757 | 0.542 | 0.78 | 0.702 |

| AUC | LASSO | LR | MLP | RF | SVM_W | XGB_W |
|---|---|---|---|---|---|---|
| Z_Original | 0.641 | 0.643 | 0.685 | 0.625 | 0.694 | 0.617 |
| Z_Raw | 0.923 | **0.935*** | 0.915 | 0.808 | 0.895 | 0.875 |
| Z_Binary | 0.835 | 0.827 | 0.815 | 0.815 | 0.853 | 0.825 |
| Z_NICG | 0.821 | 0.895* | 0.925 | 0.764 | **0.907** | 0.786 |
| Z_NPN | 0.875 | **0.915*** | 0.919 | 0.815 | 0.899 | 0.825 |
| Z_QN | 0.861 | 0.899 | 0.871 | 0.81 | 0.893 | 0.837 |
| Z_QNZ | 0.885 | 0.883* | 0.857 | 0.839 | 0.859 | 0.839 |

Note: * indicates p<0.05 when comparing with classification performance of the same machine learning model and normalization methods using data group A (all data). No markers indicate no statistical differences.

Supplementary Table 3. Performance metrics of internal testing obtained by models trained on TCGA data with molecular and 3 clinical features (age, gender and TMB) (Data grouping C)

| Balanced Accuracy | LASSO | LR | MLP | RF | SVM_W | XGB_W |
|---|---|---|---|---|---|---|
| Z_Original | 0.607 | 0.589*** | 0.595 | 0.5 | 0.599 | 0.56 |
| Z_Raw | 0.749 | **0.853** | **0.833** | 0.575 | 0.779 | 0.682 |
| Z_Binary | 0.688 | 0.74 | 0.702 | 0.538 | 0.772 | 0.684 |
| Z_NICG | 0.553 | **0.807** | 0.758 | 0.538 | 0.783 | 0.69 |
| Z_NPN | 0.636 | 0.786 | **0.824** | 0.55 | **0.802** | 0.673 |
| Z_QN | 0.681 | 0.766 | 0.749 | 0.583 | 0.754 | 0.718 |
| Z_QNZ | 0.693 | 0.766 | 0.76 | 0.558 | 0.748 | 0.712 |
| | | | | | | |
| AUC | LASSO | LR | MLP | RF | SVM_W | XGB_W |
| Z_Original | 0.643 | 0.675 | 0.687 | 0.588* | 0.606 | 0.687 |
| Z_Raw | 0.867 | **0.885** | **0.871** | 0.799 | 0.849 | 0.812 |
| Z_Binary | 0.789 | 0.825 | 0.787 | 0.787 | 0.848 | 0.789 |
| Z_NICG | 0.792 | 0.837 | 0.858 | 0.752 | **0.867** | 0.815 |
| Z_NPN | 0.848 | **0.871** | **0.882** | 0.814 | **0.888** | 0.794 |
| Z_QN | 0.845 | 0.844 | 0.827 | 0.777 | 0.849 | 0.806 |
| Z_QNZ | 0.844 | 0.84 | 0.854 | 0.76 | 0.852 | 0.827 |

Note: * and *** indicates p<0.05 and p<0.001, respectively, when comparing with classification performance of the same machine learning model and normalization methods using data group A (all data). No markers indicate no statistical differences.

Supplementary Table 4. P-values of t- test between data grouping B and data grouping C for performance metrics of internal testing obtained by models trained on TCGA data

| Balanced Accuracy | LASSO | LR | MLP | RF | SVM_W | XGB_W |
|---|---|---|---|---|---|---|
| Z_Original | 0.537 | 0.343 | 0.227 | 0.375 | 0.746 | 0.593 |
| Z_Raw | 0.915 | 0.887 | 0.211 | 0.063 | 0.276 | 0.475 |
| Z_Binary | 0.119 | 0.582 | 0.051 | 0.687 | 0.510 | 0.365 |
| Z_NICG | 0.889 | 0.270 | 0.572 | 0.304 | 0.886 | 0.625 |
| Z_NPN | 0.528 | 0.096 | 0.331 | 0.412 | 0.724 | 0.291 |
| Z_QN | 0.063 | 0.790 | 0.527 | 0.543 | 0.024 | 0.289 |
| Z_QNZ | 0.680 | 0.213 | 0.543 | 0.412 | 0.981 | 0.283 |

| AUC | LASSO | LR | MLP | RF | SVM_W | XGB_W |
|---|---|---|---|---|---|---|
| Z_Original | 0.252 | 0.153 | 0.669 | 0.979 | 0.817 | 0.117 |
| Z_Raw | 0.735 | 0.092 | 0.211 | 0.055 | 0.138 | 0.464 |
| Z_Binary | 0.458 | 0.940 | 0.752 | 0.265 | 0.070 | 0.365 |
| Z_NICG | 0.338 | 0.167 | 0.572 | 0.520 | 0.354 | 0.656 |
| Z_NPN | 0.528 | 0.096 | 0.570 | 0.064 | 0.409 | 0.291 |
| Z_QN | 0.582 | 0.654 | 0.527 | 0.543 | 0.189 | 0.289 |
| Z_QNZ | 0.198 | 0.213 | 0.989 | 0.412 | 0.981 | 0.190 |

Supplementary Table 5. Performance metrics of external testing obtained by predicting on ONCOSG data with models trained on TCGA data (including molecular and 4 clinical features) (Data grouping A)

| Balanced Accuracy | LASSO | LR | MLP | RF | SVM_W | XGB_W |
|---|---|---|---|---|---|---|
| Z_Original | 0.506 | 0.507 | 0.520 | 0.508 | 0.527 | 0.539 |
| Z_Raw | 0.553 | 0.555 | 0.574 | 0.553 | **0.608** | 0.611 |
| Z_Binary | 0.520 | 0.524 | **0.589** | 0.524 | **0.645** | 0.618 |
| Z_NICG | 0.557 | 0.552 | 0.561 | 0.536 | 0.598 | 0.583 |
| Z_NPN | 0.547 | 0.532 | 0.544 | 0.539 | 0.580 | 0.591 |
| Z_QN | 0.538 | 0.534 | 0.562 | 0.535 | **0.597** | 0.591 |
| Z_QNZ | 0.530 | 0.534 | 0.557 | 0.530 | **0.589** | 0.602 |

| AUC | LASSO | LR | MLP | RF | SVM_W | XGB_W |
|---|---|---|---|---|---|---|
| Z_Original | 0.620 | 0.639 | 0.525 | 0.648 | 0.560 | 0.574 |
| Z_Raw | 0.680 | **0.676** | 0.599 | 0.670 | 0.614 | 0.658 |
| Z_Binary | 0.685 | **0.677** | 0.632 | **0.690** | **0.654** | 0.666 |
| Z_NICG | 0.593 | 0.608 | 0.596 | 0.602 | 0.654 | 0.647 |
| Z_NPN | 0.653 | 0.658 | 0.602 | 0.662 | 0.627 | 0.649 |
| Z_QN | 0.679 | **0.677** | 0.633 | **0.678** | 0.613 | 0.681 |
| Z_QNZ | 0.668 | 0.666 | 0.612 | 0.665 | 0.613 | 0.667 |

Supplementary Table 6. Performance metrics of external testing obtained by predicting on ONCOSG data with models trained on TCGA data (including molecular features) (Data grouping B)

| Balanced Accuracy | LASSO | LR | MLP | RF | SVM_W | XGB_W |
|---|---|---|---|---|---|---|
| Z_Original | 0.539 | 0.513 | 0.527 | 0.515 | 0.586 | 0.532 |
| Z_Raw | 0.613 | 0.566 | 0.569 | 0.557 | **0.630** | 0.608 |
| Z_Binary | 0.500* | 0.500* | 0.500* | 0.500* | 0.500* | 0.500* |
| Z_NICG | 0.537 | **0.657** | 0.579 | 0.559 | **0.630** | 0.604 |
| Z_NPN | 0.578 | 0.634 | 0.529 | 0.541 | 0.621** | 0.578 |
| Z_QN | 0.549 | 0.583 | 0.550 | 0.537 | 0.610** | 0.590 |
| Z_QNZ | 0.557 | 0.577 | 0.566 | 0.529 | **0.628** | 0.583 |
| | | | | | | |
| AUC | LASSO | LR | MLP | RF | SVM_W | XGB_W |
| Z_Original | 0.553** | 0.531*** | 0.537 | 0.629 | 0.601 | 0.584 |
| Z_Raw | 0.633 | 0.564 | 0.567 | **0.701** | 0.661* | 0.655 |
| Z_Binary | 0.500*** | 0.500** | 0.500** | 0.500** | 0.500*** | 0.500** |
| Z_NICG | 0.601 | **0.687** | 0.589 | **0.686** | **0.684** | 0.647 |
| Z_NPN | 0.645 | 0.647 | 0.634 | 0.662 | 0.643 | 0.608 |
| Z_QN | 0.630 | 0.620 | 0.625 | **0.667** | 0.651 | 0.650 |
| Z_QNZ | 0.626 | 0.622 | 0.618 | **0.684** | 0.642 | 0.622 |

Note: *, ** and *** indicates p<0.05, p<0.01 and p<0.001, respectively, when comparing with classification performance of the same machine learning model and normalization methods using data group A (all data). No markers indicate no statistical differences.

Supplementary Table 7. Performance metrics of external testing obtained by predicting on ONCOSG data with models trained on TCGA data (including molecular and 3 clinical features) (Data grouping C)

| Balanced Accuracy | LASSO | LR | MLP | RF | SVM_W | XGB_W |
|---|---|---|---|---|---|---|
| Z_Original | 0.537 | 0.539 | 0.524 | 0.518* | 0.555 | 0.540 |
| Z_Raw | 0.610 | 0.580** | 0.560 | 0.550 | **0.644** | 0.580 |
| Z_Binary | 0.611 | 0.601 | 0.591 | 0.520 | **0.654** | 0.590 |
| Z_NICG | 0.575 | **0.630** | 0.573 | 0.532 | 0.606 | 0.580 |
| Z_NPN | 0.565 | **0.636** | 0.551 | 0.542 | 0.603 | 0.584 |
| Z_QN | 0.574 | 0.611 | 0.570 | 0.543 | 0.580 | 0.581 |
| Z_QNZ | 0.582 | **0.629** | 0.545 | 0.547 | 0.604 | 0.583 |
| | | | | | | |
| AUC | LASSO | LR | MLP | RF | SVM_W | XGB_W |
| Z_Original | 0.577* | 0.552* | 0.556 | 0.624 | 0.567 | 0.578 |
| Z_Raw | 0.625 | 0.585* | 0.567 | **0.670** | 0.637 | 0.631 |
| Z_Binary | **0.666** | 0.647 | 0.637 | 0.656 | **0.665** | 0.635 |
| Z_NICG | 0.579 | **0.662** | 0.585 | 0.589 | **0.673** | 0.622 |
| Z_NPN | 0.634 | 0.642 | 0.600 | 0.652 | 0.616 | 0.636 |
| Z_QN | 0.613 | 0.650 | 0.622 | 0.659 | 0.624 | 0.643 |
| Z_QNZ | 0.627 | 0.655 | 0.610 | **0.672** | 0.603 | 0.628 |

Note: *, ** and *** indicates p<0.05, p<0.01 and p<0.001, respectively, when comparing with classification performance of the same machine learning model and normalization methods using data group A (all data). No markers indicate no statistical differences.

Supplementary Table 8. P-values of t- test between data grouping B and data grouping C for performance metrics of external testing obtained by predicting on ONCOSG data with models trained on TCGA data

| Balanced Accuracy | LASSO | LR | MLP | RF | SVM_W | XGB_W |
|---|---|---|---|---|---|---|
| Z_Original | 0.252 | 0.153 | 0.669 | 0.979 | 0.817 | 0.117 |
| Z_Raw | 0.837 | 0.794 | 0.710 | 0.293 | 0.709 | 0.826 |
| Z_Binary | 0.002 | 0.003 | 0.010 | 0.000 | 0.001 | 0.252 |
| Z_NICG | 0.029 | 0.865 | 0.419 | 0.247 | 0.624 | 0.288 |
| Z_NPN | 0.976 | 0.798 | 0.261 | 0.001 | 0.007 | 0.221 |
| Z_QN | 0.917 | 0.327 | 0.696 | 0.062 | 0.061 | 0.644 |
| Z_QNZ | 0.602 | 0.075 | 0.269 | 0.459 | 0.537 | 0.626 |
| | | | | | | |
| AUC | LASSO | LR | MLP | RF | SVM_W | XGB_W |
| Z_Original | 0.144 | 0.003 | 0.319 | 0.926 | 0.817 | 0.754 |
| Z_Raw | 0.900 | 0.901 | 0.837 | 0.224 | 0.289 | 0.257 |
| Z_Binary | 0.001 | 0.003 | 0.034 | 0.000 | 0.006 | 0.002 |
| Z_NICG | 0.398 | 0.904 | 0.620 | 0.441 | 0.066 | 0.425 |
| Z_NPN | 0.478 | 0.518 | 0.867 | 0.007 | 0.594 | 0.570 |
| Z_QN | 0.599 | 0.195 | 0.980 | 0.546 | 0.524 | 0.517 |
| Z_QNZ | 0.522 | 0.062 | 0.898 | 0.742 | 0.522 | 0.927 |

Supplementary Table 9. Performance metrics of internal testing obtained by models trained on ONCOSG data with molecular and 4 clinical features (age, gender TMB and tumor stage) (Data grouping A)

| Balanced Accuracy | LASSO | LR | MLP | RF | SVM_W | XGB_W |
|---|---|---|---|---|---|---|
| Z_Original | 0.652 | 0.688 | 0.663 | 0.585 | 0.717 | 0.642 |
| Z_Raw | **0.919** | 0.881 | **0.935** | 0.727 | **0.873** | 0.788 |
| Z_Binary | 0.785 | 0.823 | 0.821 | 0.667 | 0.812 | 0.744 |
| Z_NICG | **0.904** | **0.904** | 0.854 | 0.835 | **0.921** | 0.794 |
| Z_NPN | 0.762 | 0.815 | 0.804 | 0.685 | 0.783 | 0.787 |
| Z_QN | 0.846 | 0.865 | 0.819 | 0.817 | 0.858 | 0.781 |
| Z_QNZ | 0.835 | 0.856 | 0.792 | 0.867 | 0.829 | 0.800 |
| | | | | | | |
| AUC | LASSO | LR | MLP | RF | SVM_W | XGB_W |
| Z_Original | 0.635 | 0.588 | 0.776 | 0.788 | 0.765 | 0.824 |
| Z_Raw | **0.929** | 0.894 | **0.953** | 0.859 | 0.871 | 0.859 |
| Z_Binary | 0.883 | 0.871 | 0.871 | 0.835 | 0.800 | 0.847 |
| Z_NICG | 0.906 | **0.906** | **0.906** | **0.906** | 0.906 | 0.871 |
| Z_NPN | 0.859 | 0.835 | 0.871 | 0.835 | 0.800 | 0.859 |
| Z_QN | 0.894 | 0.859 | 0.882 | **0.906** | 0.835 | 0.871 |
| Z_QNZ | 0.906 | 0.859 | 0.894 | 0.929 | 0.847 | 0.906 |

Supplementary Table 10. Performance metrics of internal testing obtained by models trained on ONCOSG data with molecular features (Data grouping B)

| Balanced Accuracy | LASSO | LR | MLP | RF | SVM_W | XGB_W |
|---|---|---|---|---|---|---|
| Z_Original | 0.683 | 0.640 | 0.648* | 0.635 | 0.742 | 0.660 |
| Z_Raw | **0.904** | 0.929 | 0.894 | 0.785 | 0.863 | 0.837 |
| Z_Binary | 0.760 | 0.860 | 0.823 | 0.669 | 0.833 | 0.894 |
| Z_NICG | 0.800 | **0.938*** | 0.877* | 0.810 | **0.977** | 0.867 |
| Z_NPN | 0.710** | **0.962** | **0.917** | 0.787 | **0.969** | 0.825 |
| Z_QN | 0.742* | 0.885 | 0.854 | 0.769 | 0.848 | 0.869 |
| Z_QNZ | 0.727 | 0.962 | 0.892 | 0.760 | 0.873 | 0.835 |

| AUC | LASSO | LR | MLP | RF | SVM_W | XGB_W |
|---|---|---|---|---|---|---|
| Z_Original | 0.647 | 0.741 | 0.765* | 0.812 | 0.765 | 0.824 |
| Z_Raw | **0.992** | 0.985 | 0.969 | 0.977 | **0.988** | 0.977 |
| Z_Binary | 0.900 | 0.908 | 0.900 | 0.896 | 0.938 | 0.942 |
| Z_NICG | 0.888 | 0.992* | **0.962** | **0.950** | **1.000*** | 0.923 |
| Z_NPN | 0.965 | 0.985 | **0.985** | 0.915 | **0.981** | 0.946 |
| Z_QN | 0.877 | 0.935 | 0.954 | 0.927 | 0.938 | 0.950 |
| Z_QNZ | 0.931 | 0.985 | **0.992** | 0.965 | **0.985**** | 0.935 |

Note: *, ** and *** indicates p<0.05, p<0.01 and p<0.001, respectively, when comparing with classification performance of the same machine learning model and normalization methods using data group A (all data). No markers indicate no statistical differences.

Supplementary Table 11. Performance metrics of internal testing obtained by models trained on ONCOSG data with molecular and 3 clinical features (age, gender and TMB) (Data grouping C)

| Balanced Accuracy | LASSO | LR | MLP | RF | SVM_W | XGB_W |
|---|---|---|---|---|---|---|
| Z_Original | 0.596 | 0.617 | 0.687 | 0.585 | 0.694 | 0.637 |
| Z_Raw | 0.885 | 0.900 | 0.871 | 0.710 | 0.846 | 0.769 |
| Z_Binary | 0.835 | 0.877 | 0.835 | 0.725 | 0.833 | 0.802 |
| Z_NICG | 0.885 | **0.923** | **0.927** | 0.810 | **0.904** | 0.844 |
| Z_NPN | 0.837 | 0.848 | 0.802 | 0.712 | 0.850 | 0.796 |
| Z_QN | 0.854 | 0.825 | 0.796 | 0.792 | 0.833 | 0.835 |
| Z_QNZ | 0.787 | 0.865 | 0.754 | 0.867 | 0.835 | 0.885 |

| AUC | LASSO | LR | MLP | RF | SVM_W | XGB_W |
|---|---|---|---|---|---|---|
| Z_Original | 0.727 | 0.735 | 0.731 | 0.792* | 0.727* | 0.765 |
| Z_Raw | **0.927** | 0.931 | **0.942** | 0.900 | 0.915 | 0.881 |
| Z_Binary | 0.885 | 0.888 | 0.927 | 0.908 | 0.908 | 0.892 |
| Z_NICG | **0.962** | **0.942** | 0.912 | 0.919 | **0.954** | 0.906 |
| Z_NPN | **0.938** | **0.938** | 0.896 | 0.915 | 0.912 | 0.935 |
| Z_QN | 0.904 | 0.900 | 0.892 | 0.912 | 0.900 | 0.965 |
| Z_QNZ | 0.912 | 0.908 | 0.885 | 0.942 | 0.896 | 0.912 |

Note: *, ** and *** indicates p<0.05, p<0.01 and p<0.001, respectively, when comparing with classification performance of the same machine learning model and normalization methods using data group A (all data). No markers indicate no statistical differences.

Supplementary Table 12. P-values of t- test between data grouping B and data grouping C for performance metrics of internal testing obtained by models trained on ONCOSG data

| Balanced Accuracy | LASSO | LR | MLP | RF | SVM_W | XGB_W |
|---|---|---|---|---|---|---|
| Z_Original | 0.207 | 0.683 | 0.004 | 0.637 | 0.319 | 0.578 |
| Z_Raw | 0.596 | 0.035 | 0.306 | 0.135 | 0.298 | 0.706 |
| Z_Binary | 0.961 | 0.948 | 0.494 | 0.679 | 0.356 | 0.242 |
| Z_NICG | 0.953 | 0.030 | 0.086 | 0.194 | 0.073 | 0.864 |
| Z_NPN | 0.097 | 0.190 | 0.957 | 0.136 | 0.132 | 0.947 |
| Z_QN | 0.010 | 0.422 | 0.242 | 0.804 | 0.535 | 0.737 |
| Z_QNZ | 0.132 | 0.020 | 0.084 | 0.367 | 0.055 | 0.722 |
| | | | | | | |
| AUC | LASSO | LR | MLP | RF | SVM_W | XGB_W |
| Z_Original | 0.621 | 0.609 | 0.021 | 0.071 | 0.308 | 0.005 |
| Z_Raw | 0.056 | 0.066 | 0.495 | 0.061 | 0.119 | 0.149 |
| Z_Binary | 0.778 | 0.768 | 0.618 | 0.759 | 0.473 | 0.319 |
| Z_NICG | 0.244 | 0.206 | 0.155 | 1.000 | 0.158 | 0.864 |
| Z_NPN | 0.051 | 0.190 | 0.488 | 0.531 | 0.033 | 0.709 |
| Z_QN | 0.587 | 0.389 | 0.065 | 0.772 | 0.483 | 0.737 |
| Z_QNZ | 0.636 | 0.031 | 0.042 | 0.393 | 0.096 | 0.722 |

Supplementary Table 13. Performance metrics of external testing obtained by predicting on TCGA data with models trained on ONCOSG data (including molecular and 4 clinical features) (Data grouping A)

| Balanced Accuracy | LASSO | LR | MLP | RF | SVM_W | XGB_W |
|---|---|---|---|---|---|---|
| Z_Original | 0.529 | 0.530 | 0.526 | 0.506 | **0.555** | 0.523 |
| Z_Raw | 0.529 | 0.530 | 0.531 | 0.521 | **0.556** | 0.541 |
| Z_Binary | 0.553 | **0.556** | 0.549 | 0.526 | **0.554** | 0.542 |
| Z_NICG | 0.543 | **0.554** | 0.536 | 0.523 | 0.533 | 0.547 |
| Z_NPN | 0.529 | 0.541 | 0.533 | 0.519 | 0.540 | 0.531 |
| Z_QN | 0.530 | **0.554** | 0.532 | 0.528 | 0.552 | 0.537 |
| Z_QNZ | 0.525 | 0.543 | 0.526 | 0.522 | 0.537 | 0.542 |

| AUC | LASSO | LR | MLP | RF | SVM_W | XGB_W |
|---|---|---|---|---|---|---|
| Z_Original | 0.550 | 0.557 | 0.570 | 0.566 | 0.585 | 0.550 |
| Z_Raw | 0.549 | 0.557 | 0.570 | 0.566 | 0.585 | 0.551 |
| Z_Binary | 0.586 | **0.579** | 0.563 | **0.593** | 0.590 | 0.561 |
| Z_NICG | **0.574** | 0.576 | 0.535 | 0.586 | 0.546 | 0.575 |
| Z_NPN | 0.567 | 0.561 | 0.541 | **0.587** | 0.554 | 0.566 |
| Z_QN | **0.574** | 0.574 | 0.564 | 0.572 | 0.552 | 0.565 |
| Z_QNZ | 0.566 | 0.564 | 0.579 | 0.570 | 0.546 | 0.546 |

Supplementary Table 14. Performance metrics of external testing obtained by predicting on TCGA data with models trained on ONCOSG data (including molecular features) (Data grouping B)

| Balanced Accuracy | LASSO | LR | MLP | RF | SVM_W | XGB_W |
|---|---|---|---|---|---|---|
| Z_Original | 0.535 | 0.537 | 0.527 | 0.504 | 0.532** | 0.519 |
| Z_Raw | 0.535 | 0.545** | 0.537 | 0.519 | **0.555*** | 0.556 |
| Z_Binary | 0.500** | 0.500* | 0.500 | 0.500* | 0.500* | 0.500** |
| Z_NICG | 0.510 | 0.544 | 0.534 | 0.532*** | 0.550 | **0.564** |
| Z_NPN | 0.531 | 0.538 | 0.532* | 0.521 | **0.552*** | 0.540 |
| Z_QN | 0.529 | 0.545 | 0.534 | 0.523 | **0.571*** | 0.553 |
| Z_QNZ | 0.528 | **0.548** | 0.536 | 0.523 | **0.563**** | 0.536 |

| AUC | LASSO | LR | MLP | RF | SVM_W | XGB_W |
|---|---|---|---|---|---|---|
| Z_Original | 0.535 | 0.537 | 0.527 | 0.504 | 0.532 | 0.519 |
| Z_Raw | 0.585 | 0.558 | 0.559 | 0.574 | **0.608** | 0.572 |
| Z_Binary | 0.507*** | 0.507*** | **0.632** | 0.500** | 0.500** | 0.500* |
| Z_NICG | 0.553 | 0.562 | 0.520 | 0.584 | **0.612** | 0.589 |
| Z_NPN | 0.557 | 0.557 | 0.547 | 0.617 | **0.596*** | 0.583 |
| Z_QN | 0.573 | 0.578 | 0.587 | 0.582 | **0.599*** | 0.582 |
| Z_QNZ | 0.570 | **0.586** | 0.587 | 0.591 | **0.596*** | 0.552 |

Note: *, ** and *** indicates $p<0.05$, $p<0.01$ and $p<0.001$, respectively, when comparing with classification performance of the same machine learning model and normalization methods using data group A (all data). No markers indicate no statistical differences.

Supplementary Table 15. Performance metrics of external testing obtained by predicting on TCGA data with models trained on ONCOSG data (including molecular and 3 clinical features) (Data grouping C)

| Balanced Accuracy | LASSO | LR | MLP | RF | SVM_W | XGB_W |
|---|---|---|---|---|---|---|
| Z_Original | 0.522 | 0.530 | 0.530 | 0.507 | **0.550** | 0.521 |
| Z_Raw | 0.529 | 0.530 | 0.536 | 0.522 | **0.554** | 0.540 |
| Z_Binary | **0.558** | 0.546 | **0.557** | 0.527 | **0.569*** | 0.535 |
| Z_NICG | 0.528 | 0.551 | 0.524 | 0.532 | 0.536 | 0.534 |
| Z_NPN | 0.531 | 0.537 | 0.531* | 0.514 | 0.535 | 0.537* |
| Z_QN | 0.524 | 0.537 | 0.525 | 0.526 | 0.545 | 0.541 |
| Z_QNZ | 0.527 | 0.537 | 0.532 | 0.533 | **0.554** | 0.540 |
|  |  |  |  |  |  |  |
| AUC | LASSO | LR | MLP | RF | SVM_W | XGB_W |
| Z_Original | 0.554 | 0.550 | 0.537* | 0.555 | 0.576 | 0.554 |
| Z_Raw | 0.554 | 0.550 | 0.562 | 0.562 | 0.576 | 0.554 |
| Z_Binary | **0.577** | 0.570 | 0.567 | **0.601** | **0.595** | 0.561 |
| Z_NICG | 0.562 | 0.565 | 0.533 | 0.577 | 0.554 | **0.584** |
| Z_NPN | 0.544 | 0.545 | 0.539 | **0.584** | 0.546 | 0.576 |
| Z_QN | **0.577** | **0.577** | 0.572 | **0.582** | 0.552 | 0.554 |
| Z_QNZ | 0.566 | 0.590 | 0.562 | 0.570 | 0.552 | 0.574 |

Note: *, ** and *** indicates p<0.05, p<0.01 and p<0.001, respectively, when comparing with classification performance of the same machine learning model and normalization methods using data group A (all data). No markers indicate no statistical differences.

Supplementary Table 16. P-values of t-test between data grouping B and data grouping C for performance metrics of external testing obtained by predicting on TCGA data with models trained on ONCOSG data

| Balanced Accuracy | LASSO | LR | MLP | RF | SVM_W | XGB_W |
|---|---|---|---|---|---|---|
| Z_Original | 0.568 | 0.328 | 0.197 | 0.537 | 0.001 | 0.897 |
| Z_Raw | 0.874 | 0.021 | 0.689 | 0.745 | 0.105 | 0.804 |
| Z_Binary | 0.026 | 0.029 | 0.059 | 0.044 | 0.001 | 0.014 |
| Z_NICG | 0.563 | 0.009 | 0.907 | 0.001 | 0.040 | 0.573 |
| Z_NPN | 0.550 | 0.895 | 0.389 | 0.352 | 0.009 | 0.007 |
| Z_QN | 0.058 | 0.985 | 0.237 | 0.561 | 0.012 | 0.964 |
| Z_QNZ | 0.284 | 0.449 | 0.583 | 0.239 | 0.117 | 0.953 |
|  |  |  |  |  |  |  |
| AUC | LASSO | LR | MLP | RF | SVM_W | XGB_W |
| Z_Original | 0.286 | 0.891 | 0.828 | 0.172 | 0.300 | 0.227 |
| Z_Raw | 0.149 | 0.740 | 0.921 | 0.515 | 0.061 | 0.435 |
| Z_Binary | 0.045 | 0.001 | 0.222 | 0.005 | 0.001 | 0.014 |
| Z_NICG | 0.432 | 0.374 | 0.730 | 0.311 | 0.050 | 0.719 |

| Z_NPN | 0.387 | 0.807 | 0.620 | 0.761 | 0.060 | 0.789 |
| Z_QN  | 0.808 | 0.922 | 0.651 | 0.997 | 0.002 | 0.071 |
| Z_QNZ | 0.824 | 0.449 | 0.337 | 0.313 | 0.039 | 0.250 |

Supplementary Table 17. Comparison of the best performances (balanced accuracy) in data grouping A, B and C based on the model trained on TCGA data.

|  | Internal testing | | | | | | External independent testing | | | | | |
|---|---|---|---|---|---|---|---|---|---|---|---|---|
|  | Data grouping | | | P-value | | | Data grouping | | | P-value | | |
|  | A | B | C | A v. B | A v. C | B v. C | A | B | C | A v. B | A v. C | B v. C |
| Balanced accuracy | 0.814+0.010 | 0.848+0.001 | 0.853+0.011 | 0.500 | 0.564 | 0.923 | 0.645+0.003 | 0.657+0.001 | 0.654+0.001 | 0.685 | 0.761 | 0.885 |
| AUC | 0.888±0.023 | 0.925±0.019 | 0.885±0.008 | 0.697 | 0.971 | 0.603 | 0.654±0.002 | 0.687±0.001 | 0.665±0.002 | 0.219 | 0.708 | 0.398 |
| Accuracy | 0.821±0.006 | 0.890±0.001 | 0.910±0.005 | 0.121 | 0.09 | 0.586 | 0.645±0.003 | 0.657±0.001 | 0.654±0.001 | 0.685 | 0.761 | 0.885 |
| DEG | 78 | 1430 | 996 | | | | 161 | 176 | 816 | | | |
| NDEG | 62 | 120 | 120 | | | | 120 | 120 | 120 | | | |
| Normalization method | Z_Raw | Z_NPN | Z_Raw | | | | Z_Binary | Z_NICG | Z_Binary | | | |
| Classification model | SVM_W | MLP | LR | | | | SVM_W | LR | SVM_W | | | |

All data are shown in mean ± standard deviation. Data grouping: A, molecular data plus 4 clinical features; B, molecular data alone; C, molecular data plus 3 clinical features. All experiments were repeated 15 times. V., versus.

Supplementary Table 18. Comparison of the best performances (balanced accuracy) of data grouping A, B and C based on the model trained on ONCOSG data.

|  | Internal testing | | | | | | External independent testing | | | | | |
|---|---|---|---|---|---|---|---|---|---|---|---|---|
|  | Data grouping | | | P-value | | | Data grouping | | | P-value | | |
|  | A | B | C | A v. B | A v. C | B v. C | A | B | C | A v. B | A v. C | B v. C |
| Balanced accuracy | 0.935 ± 0.004 | 0.977± 0.000 | 0.927± 0.003 | 0.212 | 0.836 | 0.111 | 0.556± 0.000* | 0.571± 0.000* | 0.569± 0.000* | 0.0001 ### | 0.0001 ### | 0.013 # |
| AUC | 0.977 ± 0.001 | 1.000± 0.000* | 0.912± 0.010 | 0.179 | 0.227 | 0.121 | 0.579± 0.000* | 0.599± 0.000* | 0.595± 0.000* | 0.0001 ### | 0.0001 ### | 0.0001 ### |
| Accuracy | 0.953 ± 0.002 | 0.965± 0.001 | 0.941± 0.002 | 0.639 | 0.683 | 0.359 | 0.556± 0.000* | 0.571± 0.000* | 0.569± 0.000* | 0.0001 ### | 0.0001 ### | 0.013 # |
| DEG | 534 | 2382 | 2070 | | | | 617 | 2382 | 2960 | | | |
| NDEG | 230 | 65 | 65 | | | | 1729 | 230 | 562 | | | |
| Normalization method | Z_Raw | Z_NICG | Z_NICG | | | | Z_Binary | Z_QN | Z_Binary | | | |
| Classification model | MLP | SVM_W | MLP | | | | LR | SVM_W | SVM_W | | | |

All data are shown in mean ± standard deviation. Data grouping: A, molecular data plus 4 clinical features; B, molecular data alone; C, molecular data plus 3 clinical features. All experiments were repeated 15 times. * The SD was less than 0.00001. #, 0.01≤P-value<0.05. ##, 0.001≤P-value<0.01. ###, P-value<0.001.